\documentclass[12pt]{iopart}
\usepackage{iopams}
\usepackage{graphicx}
\usepackage{bm}
\newcommand{\Dq}{\mathcal{D}}
\newcommand{\Gq}{\mathcal{G}}
\newcommand{\Hq}{\mathcal{H}}
\newcommand{\Pq}{\mathcal{P}}
\expandafter\let\csname equation*\endcsname\relax
\expandafter\let\csname endequation*\endcsname\relax
\usepackage{amsmath}
\newtheorem{theorem}{Theorem}
\newenvironment{proof}{\noindent{\it Proof.}\ }{\hfill$\square$\par\medskip}

\begin{document}

\title[Retarded relativistic Bethe sum rules]{Sum rules for the relativistic and retarded Bethe
approximation in electron--neutral-atom collisions}

\author{Colm T Whelan}

\address{Department of Physics, Old Dominion University, Norfolk, Virginia 23529, USA}
\ead{cwhelan@odu.edu}

\begin{abstract}
For high partial waves the reactance matrix for electron--atom collisions may be replaced by the
Bethe approximation, and for optically allowed transitions the slowly convergent sum over partial
waves can then be performed exactly by means of the sum rules of Burgess. For neutral targets the
Bethe integrals and the sum rules take a particularly simple form in terms of Legendre functions of
the second kind of a single argument $\chi=(k^2+k'^2)/2kk'$. We show that this structure survives
in a fully relativistic description. For a projectile described by free Dirac waves interacting
through the retarded M{\o}ller interaction, the Bethe integrals of every multipole order are
obtained from the static ones by the replacement $\chi\to\chi_q=(k^2+k'^2-q^2)/2kk'$, where $q$ is
the energy transfer divided by $c$. For dipole transitions we give closed-form sum rules over all
partial waves above any $l_0\ge2$ for the Coulomb (charge-density) interaction, for the magnetic
dipole interaction and for the long-range transverse electric interaction, including its
interference with the Coulomb term. In each case the sum reduces to a quadratic form in
$Q_{l_0-1}(\chi_q)$ and $Q_{l_0}(\chi_q)$ with coefficients rational in $l_0$. As the impact energy
increases $\chi_q\to1$ and the partial-wave sums converge ever more slowly, so that the sum rules
become most useful precisely in the relativistic regime.
\end{abstract}


\maketitle

\section{Introduction}

In partial-wave calculations of electron--atom collisions at high impact energies the sum over the
angular momentum of the projectile converges slowly whenever a long-range interaction is present.
For optically allowed transitions the contribution of the high partial waves is dominated by the
dipole part of the interaction between the projectile and the target, and it is common to replace
the reactance matrix for large $l$ by its Born value or, more simply, by the Bethe approximation, in
which the short-range part of the interaction is neglected \cite{seaton61,burgess70}. Burgess
\cite{burgess74} showed that within the Bethe approximation the sum over all partial waves above a
given $l_0$ can be carried out exactly for dipole transitions in a Coulomb field, the result
depending only on Coulomb integrals at $l_0$; the derivation rests on recurrence relations for the
Coulomb functions in the manner of Biedenharn \cite{biedenharn56}, and the method was applied by
Burgess and Sheorey \cite{burgess74a}. For electron--neutral-atom collisions the integrals reduce to
integrals over two spherical Bessel functions, and it was shown in \cite{whelan86} that these
integrals, and with them the Burgess sum rules, then take a particularly simple form in terms of
Legendre functions of the second kind of the single argument $\chi=(k^2+k'^2)/2kk'$, where $k$ and
$k'$ are the wave numbers of the projectile before and after the collision. The Legendre functions
are computed accurately and cheaply by backward recurrence, and the method has been implemented in
\cite{burgess87}.

At relativistic impact energies the projectile must be described by Dirac waves and its interaction
with the target by the M{\o}ller interaction, which contains, besides the Coulomb interaction
between the charge densities, the magnetic interaction between the currents and the effects of
retardation. The relativistic distorted-wave Born approximation (rDWBA) for inner-shell ionization of
heavy atoms \cite{keller94a,nakel99} is an example: there the number of partial waves required grows
with the momentum of the projectile, and the slow convergence of the partial-wave sums limits the
energies that can be reached. It is therefore natural to ask whether the Burgess sum rules
generalize.

In this paper we show that for neutral targets they do, and that the generalization is
surprisingly simple. Our main results are the following. (i) For a projectile described by free
Dirac waves and interacting through the retarded M{\o}ller interaction, the Bethe integrals of
every multipole order are obtained from the static ones of \cite{whelan86} by replacing $\chi$ by
\begin{equation}
\chi_q=\frac{k^2+k'^2-q^2}{2kk'},\qquad q=\frac{\omega}{c},
\label{eq:chiq}
\end{equation}
where $\omega$ is the energy transferred to the target (section~\ref{sec:retarded}). (ii) For dipole
transitions the sums over all partial waves above any $l_0\ge2$ of the Coulomb, magnetic and
long-range transverse electric terms, including the interference between the Coulomb and the
transverse terms, can be carried out in closed form; each sum is a quadratic form in
$Q_{l_0-1}(\chi_q)$ and $Q_{l_0}(\chi_q)$ with coefficients rational in $l_0$
(section~\ref{sec:sumrules}). (iii) Retardation drives $\chi_q$ towards unity as the energy
increases, which lengthens the partial-wave tail dramatically; the sum rules are therefore most
useful exactly where they are most needed (section~\ref{sec:numerics}). Atomic units are used
throughout.

\section{The nonrelativistic case}
\label{sec:nr}

We summarize the results of \cite{whelan86} that are needed below. For a $2^\lambda$-pole
transition of the target the Bethe approximation to the radial factor of the reactance matrix is
the product of a target factor and the integral
\begin{equation}
I_0(k,l,k',l';\lambda)=(kk')^{1/2}\int_0^\infty j_{l'}(k'r)\,j_l(kr)\,r^{-\lambda+1}\,dr ,
\label{eq:I0}
\end{equation}
where $k$, $l$ and $k'$, $l'$ are the wave number and orbital angular momentum of the projectile
before and after the collision. Only the combinations allowed by the angular factors of the
reactance matrix are required, that is $l+l'+\lambda$ even and $|l-l'|\le\lambda\le l+l'$,
and $k\ne k'$ for inelastic transitions. It was shown in \cite{whelan86} that
\begin{equation}
I_0(k,l,k',l';\lambda)=\sqrt{\frac{\pi}{kk'}}\,\frac{1}{2^\lambda\Gamma(\lambda+\frac{1}{2})}
\sum_p\mathcal{Q}_{\lambda,p}(k,l,k',l')\,Q_p(\chi),
\label{eq:betheint}
\end{equation}
where the coefficients $\mathcal{Q}_{\lambda,p}(k,l,k',l')$ are defined in \cite{seaton61} and
$Q_p(\chi)$ is a Legendre function of the second kind, with $\chi=(k^2+k'^2)/2kk'$. For $k\ne k'$
we have $\chi>1$, which avoids the singularity of the Legendre function at $\chi=1$.
The statement in \cite{whelan86} that
$I_0(k,l,k',l';0)$ vanishes for $l\ne l'$ holds, as Qiu \etal \cite{qiu07} note, only for the
values of $l$, $l'$ and $\lambda$ allowed by the selection rules of the reactance matrix.
For dipole transitions
\begin{eqnarray}
I_0(k,l,k',l-1;1)&=&\frac{1}{2(kk')^{1/2}}\left[kQ_{l-1}(\chi)-k'Q_l(\chi)\right],\nonumber\\
I_0(k,l,k',l+1;1)&=&\frac{1}{2(kk')^{1/2}}\left[k'Q_{l}(\chi)-kQ_{l+1}(\chi)\right],
\label{eq:I0dip}
\end{eqnarray}
and for $\lambda=0$, $l'=l$,
\begin{equation}
I_0(k,l,k',l;0)=\frac{Q_l(\chi)}{2(kk')^{1/2}} .
\label{eq:I0mono}
\end{equation}
The Burgess sum rule \cite{burgess74} for the dipole collision strength then becomes
\cite{whelan86}
\begin{equation}
\sum_{l=l_0}^\infty\sum_{l'=l\pm1} l_>\,I_0^2(k,l,k',l';1)=\frac{l_0k}{4k'}
\left[Q_{l_0-1}^2(\chi)-Q_{l_0}^2(\chi)\right],
\label{eq:nr}
\end{equation}
where $l_>$ is the larger of $l$ and $l'$. Because the $Q_l(\chi)$, $\chi>1$, are the minimal
solution of the Legendre recurrence
\begin{equation}
(l+1)Q_{l+1}(\chi)=(2l+1)\chi Q_l(\chi)-lQ_{l-1}(\chi),
\label{eq:rec}
\end{equation}
they are computed accurately by backward recurrence \cite{whelan86,burgess87}, whereas forward
recurrence is unstable.

\section{The relativistic Bethe approximation}
\label{sec:rel}

\subsection{Free Dirac waves}
Outside a neutral atom the continuum orbitals of the projectile are free Dirac spherical waves,
\begin{equation}
\psi_{\kappa m}=\frac{1}{r}\left(\begin{array}{c} g_\kappa(r)\,\Omega_{\kappa m}\\ \rmi f_\kappa(r)\,\Omega_{-\kappa m}
\end{array}\right),\qquad
g_\kappa=A(E)\,k^{1/2}rj_l(kr),\quad f_\kappa=S_\kappa B(E)\,k^{1/2}rj_{\bar l}(kr),
\label{eq:dirac}
\end{equation}
where $E=(c^4+k^2c^2)^{1/2}$ is the total energy, $S_\kappa$ is the sign of $\kappa$, $l$ and
$\bar l$ are the orbital angular momenta of the large and small components ($\bar l=l+1$ for
$\kappa=-(l+1)$, $j=l+\frac12$; $\bar l=l-1$ for $\kappa=l$, $j=l-\frac12$), the spinor spherical
harmonics are related by $\bm\sigma\cdot\hat{\bm r}\,\Omega_{\kappa m}=-\Omega_{-\kappa m}$, and
\begin{equation}
A(E)=\left(\frac{E+c^2}{2E}\right)^{1/2},\qquad B(E)=\left(\frac{E-c^2}{2E}\right)^{1/2}.
\end{equation}
The products of these factors that occur below are denoted
\begin{equation}
\alpha=A(E)A(E'),\qquad \beta=B(E)B(E'),\qquad \gamma_1=B(E)A(E'),\qquad \gamma_2=A(E)B(E'),
\label{eq:abg}
\end{equation}
where $E'$ is the total energy of the projectile after the collision; $\beta$, $\gamma_1$ and
$\gamma_2$ vanish in the nonrelativistic limit. The normalization reduces to that of
\cite{whelan86} in the nonrelativistic limit $A\to1$, $B\to0$. Phase shifts, which are small for the
large $l$ of interest, are omitted as in the Bethe approximation.

\subsection{The M{\o}ller interaction in the Bethe region}
The first-order interaction between the projectile and a target electron is the retarded M{\o}ller
interaction, which in the Feynman gauge is
\begin{equation}
V_{12}=\left(1-\bm\alpha_1\cdot\bm\alpha_2\right)\frac{\rme^{\rmi qr_{12}}}{r_{12}},\qquad
\frac{\rme^{\rmi qr_{12}}}{r_{12}}=\rmi q\sum_{\lambda}(2\lambda+1)\,j_\lambda(qr_<)\,h^{(1)}_\lambda(qr_>)\,
\bm C^{(\lambda)}(1)\cdot\bm C^{(\lambda)}(2),
\label{eq:moller}
\end{equation}
with $q=\omega/c$ and $\bm C^{(\lambda)}$ the renormalized spherical harmonics. In the Bethe
approximation the projectile is outside the target, $r_<=r_t$, $r_>=r$. The static limit $q\to0$
gives the familiar multipole expansion of $(1-\bm\alpha_1\cdot\bm\alpha_2)/r_{12}$, and it is
convenient to normalize the projectile and target factors so that this limit is reproduced: we
write the radial kernel of the projectile as
\begin{equation}
P_\lambda(r)=-\frac{q^{\lambda+1}}{(2\lambda-1)!!}\,y_\lambda(qr)\ \longrightarrow\ r^{-\lambda-1}\qquad(q\to0),
\label{eq:Plam}
\end{equation}
and absorb $(2\lambda-1)!!\,(2\lambda+1)\,j_\lambda(qr_t)/q^\lambda\to r_t^\lambda$ into the target
factor. That only the spherical Neumann function $y_\lambda$ of $h^{(1)}_\lambda=j_\lambda+\rmi y_\lambda$
contributes is shown in section~\ref{sec:retarded}.

For an electric dipole (E1) transition of the target three terms contribute. The Coulomb term,
from the $\lambda=1$ part of (\ref{eq:moller}), couples the charge density of the projectile through
$\bm C^{(1)}P_1(r)$. The magnetic interaction $\bm\alpha_1\cdot\bm\alpha_2\,\bm C^{(\lambda)}(1)\cdot\bm C^{(\lambda)}(2)$
can be recoupled into scalar products of rank-$\nu$ tensors $[\bm\alpha\otimes\bm C^{(\lambda)}]^{(\nu)}$
on each particle; since $\bm\alpha$ is odd under parity, the rank-1 odd-parity (E1) target
operators arise from $\lambda=0$ and $\lambda=2$. The $\lambda=0$ term,
$-\bm\alpha_1\cdot\bm\alpha_2\,P_0(r)$ with $P_0(r)=\cos(qr)/r$, is the longest-ranged of all; it is
responsible for the transverse (``relativistic rise'') contribution to the Bethe cross section and
interferes with the Coulomb term. The $\lambda=2$ term has a projectile kernel $P_2\to r^{-3}$; its
partial-wave sum converges rapidly and is best evaluated directly. For a magnetic dipole (M1)
transition of the target the relevant term is the rank-1 even-parity part of
$\bm\alpha_1\cdot\bm\alpha_2\,\bm C^{(1)}(1)\cdot\bm C^{(1)}(2)$, the magnetic dipole term proper, with
kernel $P_1(r)$.

\subsection{Radial factors}
With (\ref{eq:dirac}) the radial factors of the projectile reduce to integrals of the type
(\ref{eq:I0}) with $r^{-\lambda+1}$ replaced by $r^2P_\lambda(r)$, which we denote by
$I_q(k,l,k',l';\lambda)$. The Coulomb dipole factor is
\begin{equation}
R^{C}_{\kappa'\kappa}=\int_0^\infty\left(g_{\kappa'}g_\kappa+f_{\kappa'}f_\kappa\right)P_1\,dr
=\alpha\,I_q(k,l,k',l';1)+S_\kappa S_{\kappa'}\,\beta\,I_q(k,\bar l,k',\bar l';1);
\label{eq:RC}
\end{equation}
both integrals are of dipole type, since parity requires $l+l'$ odd and then $\bar l+\bar l'$ is also
odd. The magnetic dipole factor is \cite{grant07,johnson07}
\begin{equation}
R^{M}_{\kappa'\kappa}=\int_0^\infty\left(g_{\kappa'}f_\kappa+f_{\kappa'}g_\kappa\right)P_1\,dr
=S_\kappa\gamma_1\,I_q(k,\bar l,k',l';1)+S_{\kappa'}\gamma_2\,I_q(k,l,k',\bar l';1),
\label{eq:RM}
\end{equation}
with the parity condition $l+\bar l'$ odd. The transverse ($\lambda=0$) operator $\bm\alpha$ connects
the large component of one wave with the small component of the other, and since $\bm C^{(0)}=1$ the
orbital angular momenta must be equal; its two radial factors are
\begin{eqnarray}
R^{a}_{\kappa'\kappa}&=&\int_0^\infty g_{\kappa'}f_\kappa\,P_0\,dr=S_\kappa\gamma_1\,I_q(k,\bar l,k',\bar l;0)\qquad(l'=\bar l),\nonumber\\
R^{b}_{\kappa'\kappa}&=&\int_0^\infty f_{\kappa'}g_\kappa\,P_0\,dr=S_{\kappa'}\gamma_2\,I_q(k,l,k',l;0)\qquad(\bar l'=l).
\label{eq:Rab}
\end{eqnarray}

\section{Retarded Bethe integrals}
\label{sec:retarded}

\begin{theorem}
Let $k>k'$, $q=\omega/c$ with $\omega=E-E'$, and let $l$, $l'$, $\lambda$ satisfy the conditions of
section~\ref{sec:nr}. Then
\begin{equation}
I_q(k,l,k',l';\lambda)\equiv(kk')^{1/2}\int_0^\infty j_{l'}(k'r)\,j_l(kr)\,r^2P_\lambda(r)\,dr
=I_0(k,l,k',l';\lambda)\Big|_{\chi\to\chi_q},
\label{eq:theorem}
\end{equation}
and the corresponding integral with $j_\lambda(qr)$ in place of $y_\lambda(qr)$ vanishes.
\end{theorem}

\begin{proof}
Since $\rmd E/\rmd k=kc^2/E<c$, we have $\omega<c(k-k')$, that is $q<k-k'$, and $\chi_q>1$.
Inserting Seaton's representation \cite{seaton61} of the product of Bessel functions
\cite[eq.~(3.2)]{whelan86},
\begin{equation}
j_l(kr)\,j_{l'}(k'r)\ \to\ \sum_p \mathcal{Q}_{\lambda,p}(k,l,k',l')\int_{-1}^{1}
\frac{j_\lambda(Kr)}{K^\lambda}P_p(\mu)\,d\mu,\qquad K^2=2kk'(\chi-\mu),
\label{eq:seaton}
\end{equation}
which with
\begin{equation}
\int_0^\infty j_\lambda(Kr)\,r^{-\lambda+1}\,dr=\frac{\sqrt\pi\,K^{\lambda-2}}{2^\lambda\Gamma(\lambda+\frac12)}
\label{eq:static}
\end{equation}
and Neumann's integral $\frac12\int_{-1}^1P_p(\mu)(\chi-\mu)^{-1}d\mu=Q_p(\chi)$ gives
(\ref{eq:betheint}) in the static case, we need the radial integrals
\begin{eqnarray}
\int_0^\infty j_\lambda(Kr)\,j_\lambda(qr)\,r^2dr&=&\frac{\pi}{2K^2}\,\delta(K-q),\nonumber\\
\int_0^\infty j_\lambda(Kr)\,y_\lambda(qr)\,r^2dr&=&-\frac{(K/q)^\lambda}{q\,(K^2-q^2)}\qquad(K>q),
\label{eq:kernel}
\end{eqnarray}
the second being a Weber--Schafheitlin integral \cite[\S13.4]{watson44}, understood with a convergence
factor $\rme^{-\epsilon r}$, $\epsilon\to0$. On the range of integration $K\ge k-k'>q$, so the first
integral, which carries the delta function, does not contribute. With (\ref{eq:Plam}) the second gives
\begin{equation}
\int_0^\infty j_\lambda(Kr)\,r^2P_\lambda(r)\,dr=\frac{K^\lambda}{(2\lambda-1)!!\,(K^2-q^2)} ,
\end{equation}
which reduces to (\ref{eq:static}) when $q\to0$. The factor $K^{-\lambda}$ in (\ref{eq:seaton})
therefore leaves, in place of $K^{-2}=[2kk'(\chi-\mu)]^{-1}$ in the static case, the factor
$(K^2-q^2)^{-1}=[2kk'(\chi_q-\mu)]^{-1}$, and Neumann's integral gives $Q_p(\chi_q)$ in place of
$Q_p(\chi)$. All other factors are unchanged.
\end{proof}

In particular (\ref{eq:I0dip}) and (\ref{eq:I0mono}) hold for the retarded integrals with
$\chi\to\chi_q$. We have checked (\ref{eq:kernel}) and (\ref{eq:theorem}) for $\lambda=0$, 1, 2 and 3
by direct numerical integration, with relative differences below $10^{-6}$ that are set by the
quadrature. Retardation thus changes nothing in the structure of the Bethe integrals: it only moves
the argument of the Legendre functions from $\chi$ towards unity, since
$\chi-\chi_q=q^2/2kk'$. We note for later use the quantity
\begin{equation}
\Delta\equiv k^2+k'^2-2\chi kk' ,
\label{eq:Delta}
\end{equation}
which vanishes for $\chi$ given by its static value and equals $q^2$ for $\chi=\chi_q$.

\section{Angular factors}
\label{sec:angular}

The collision strength requires the squares of the projectile matrix elements summed over the
magnetic quantum numbers $m$, $m'$ and over the spherical components of the rank-1 operator. For the
Coulomb term this gives the square of the reduced matrix element of $C^{(1)}$ between spinor
spherical harmonics, which is the same for the large and the small components,
\begin{equation}
W^{C}_{\kappa'\kappa}=\tfrac12(2j+1)(2j'+1)\left(\begin{array}{ccc} j'&1&j\\ -\frac12&0&\frac12\end{array}\right)^2,
\end{equation}
which takes the values $(j+\frac12)(j+\frac32)/[2(j+1)]$, $(j+\frac12)/[4j(j+1)]$ and
$(j-\frac12)(j+\frac12)/(2j)$ for $j'=j+1$, $j$ and $j-1$; summed over $j$ and $j'$ at fixed $l$,
$l'$ it reduces to $l_>$. For the magnetic dipole term \cite{grant07}
\begin{equation}
W^{M}_{\kappa'\kappa}=\tfrac12(\kappa+\kappa')^2(2j+1)(2j'+1)\left(\begin{array}{ccc} j'&1&j\\ -\frac12&0&\frac12\end{array}\right)^2.
\end{equation}
For the transverse term and its interference with the Coulomb term the large and small components
enter with different angular factors, and the relative phases matter. We have therefore evaluated
the matrix elements of $\bm C^{(1)}$ and $\bm\alpha$ explicitly from the spinors (\ref{eq:dirac}),
summing over $m$, $m'$ and the spherical component, and normalized them in the same way as
$W^C$. The results are rational functions of $l$ and are listed in table~\ref{tab:ang}. With the
amplitude written as $\sum_\mu(-1)^\mu[t^C_\mu V^C_{-\mu}-t^0_\mu V^0_{-\mu}]$, where $t^C$ and
$t^0$ are the target factors, the projectile factor of the transverse term is
$V^0=\rmi(c_aR^a-c_bR^b)$ with angular coefficients $c_a$, $c_b$, and the $m$-summed bilinears are
\begin{eqnarray}
X^{CC}=W^{CC}(R^C)^2,\qquad X^{C0}=W^{Ca}R^CR^a+W^{Cb}R^CR^b,\nonumber\\
X^{00}=W^{aa}(R^a)^2+W^{ab}R^aR^b+W^{bb}(R^b)^2 .
\end{eqnarray}
The phase conventions were checked by verifying that (\ref{eq:dirac}) satisfies the free radial
Dirac equations and that $\bm\sigma\cdot\hat{\bm r}\,\Omega_{\kappa m}=-\Omega_{-\kappa m}$; the
Coulomb entries of table~\ref{tab:ang} reproduce $W^C$.

\begin{table}
\caption{\label{tab:ang}Angular coefficients for the Coulomb and $\lambda=0$ transverse dipole
terms, for projectile $\kappa\to\kappa'$; $l$ is the orbital angular momentum of the large component
of the incident wave. Missing entries vanish.}
\centering
\resizebox{\textwidth}{!}{\begin{tabular}{@{}llllllll}
\br
$\kappa$&$\kappa'$&$W^{CC}$&$W^{Ca}$&$W^{Cb}$&$W^{aa}$&$W^{ab}$&$W^{bb}$\\
\mr
$-(l+1)$&$-(l+2)$&$\frac{(l+1)(l+2)}{2l+3}$&$-\frac{2(l+1)(l+2)}{2l+3}$&&$\frac{4(l+1)(l+2)}{2l+3}$&&\\[3pt]
$-(l+1)$&$l+1$&$\frac{l+1}{(2l+1)(2l+3)}$&$\frac{l+1}{2l+3}$&$\frac{l+1}{2l+1}$&$\frac{(l+1)(2l+1)}{2l+3}$&$2(l+1)$&$\frac{(l+1)(2l+3)}{2l+1}$\\[3pt]
$-(l+1)$&$-l$&$\frac{l(l+1)}{2l+1}$&&$\frac{2l(l+1)}{2l+1}$&&&$\frac{4l(l+1)}{2l+1}$\\[3pt]
$l$&$l+1$&$\frac{l(l+1)}{2l+1}$&&$\frac{2l(l+1)}{2l+1}$&&&$\frac{4l(l+1)}{2l+1}$\\[3pt]
$l$&$-l$&$\frac{l}{(2l-1)(2l+1)}$&$-\frac{l}{2l-1}$&$-\frac{l}{2l+1}$&$\frac{l(2l+1)}{2l-1}$&$2l$&$\frac{l(2l-1)}{2l+1}$\\[3pt]
$l$&$l-1$&$\frac{l(l-1)}{2l-1}$&$-\frac{2l(l-1)}{2l-1}$&&$\frac{4l(l-1)}{2l-1}$&&\\
\br
\end{tabular}}
\end{table}

\section{Sum rules}
\label{sec:sumrules}

Let
\begin{equation}
\Sigma^{X}(l_0)=\sum_{l=l_0}^\infty\ \sum_{\kappa=-(l+1),\,l}\ \sum_{\kappa'}X_{\kappa'\kappa},
\qquad X=W^C(R^C)^2,\ W^M(R^M)^2,\ X^{C0},\ X^{00},
\end{equation}
the sum running over all $\kappa'$ allowed by angular momentum and parity. Write
$a=Q_{l_0-1}(\chi)$, $b=Q_{l_0}(\chi)$ and define the three quadratic forms
\begin{equation}
\Dq=a^2-b^2,\qquad \Gq=a^2-2\chi ab+b^2,\qquad \Hq=(\chi^2-1)\Dq+l_0\Gq ,
\label{eq:forms}
\end{equation}
and
\begin{equation}
\Pq=(2\chi^2-1)l_0^2\Dq+(\chi^2+1)l_0(a^2+b^2)-\chi^2\Dq-4\chi l_0ab .
\end{equation}
Here $\chi$ is either the static $\chi$ or the retarded $\chi_q$, and $\Delta$ is given by
(\ref{eq:Delta}); the results hold for arbitrary $\chi>1$, independently of its relation to $k$
and $k'$. Then, for $l_0\ge2$ and $k\ne k'$:

\noindent\emph{Coulomb dipole term:}
\begin{equation}
\Sigma^{C}(l_0)=\left(\alpha^2+2\chi\alpha\beta\right)S_0+\beta^2S_\beta ,
\label{eq:SC}
\end{equation}
\begin{equation}
S_0=\frac{l_0k}{4k'}\Dq+\frac{l_0^2\,\Delta}{4kk'(\chi^2-1)}\Gq,\qquad
S_\beta=\frac{l_0k}{4k'(l_0^2-1)}\Pq+\frac{l_0\,\Delta}{4kk'(\chi^2-1)}\Hq .
\end{equation}

\noindent\emph{Magnetic dipole term:}
\begin{equation}
\Sigma^{M}(l_0)=\gamma_1^2M_{11}+\gamma_1\gamma_2M_{12}+2\gamma_2^2S_0,
\label{eq:SM}
\end{equation}
\begin{eqnarray}
M_{11}&=&\frac{l_0}{4k'(l_0^2-1)}\Big\{k\big[2(\chi^2-1)l_0^2\Dq+(\chi^2+1)l_0(a^2+b^2)\nonumber\\
&&-(\chi^2-1)\Dq-4\chi l_0ab\big]+2\chi k'(l_0^2-1)\Dq\Big\}+\frac{l_0\,\Delta}{2kk'(\chi^2-1)}\Hq ,\nonumber\\
M_{12}&=&-\frac{l_0}{2k'}\left[\chi k\Dq+k'\left((2l_0-1)a^2-4\chi l_0ab+(2l_0+1)b^2\right)\right]\nonumber\\
&&-\frac{\chi l_0^2\,\Delta}{kk'(\chi^2-1)}\Gq .
\end{eqnarray}

\noindent\emph{Transverse ($\lambda=0$) term and its interference with the Coulomb term:}
\begin{eqnarray}
\Sigma^{C0}(l_0)&=&-\frac{l_0}{4k'}\left(\alpha\gamma_1-\frac{k'-2\chi k}{k}\,\beta\gamma_1\right)\Dq\nonumber\\
&&-\frac{l_0^2\,\Gq}{4kk'(\chi^2-1)}\Big[(k-\chi k')(\alpha\gamma_1+\beta\gamma_2)
\nonumber\\
&&\qquad+(\chi k-k')(\alpha\gamma_2+\beta\gamma_1)\Big],
\label{eq:SC0}\\
\Sigma^{00}(l_0)&=&\frac{l_0\left[3\gamma_1^2\Hq-(2\chi\gamma_1\gamma_2-3\gamma_2^2)\,l_0\Gq\right]}{4kk'(\chi^2-1)}.
\label{eq:S00}
\end{eqnarray}

In the static case $\Delta=0$ and $\Sigma^C$ reduces to the static relativistic sum rule; in the
nonrelativistic limit $\alpha\to1$, $\beta,\gamma_1,\gamma_2\to0$ and (\ref{eq:SC}) reduces to
(\ref{eq:nr}). The large--small interference term in the Coulomb sum rule is $2\chi$ times the
large--large term, and the magnetic $\gamma_2^2$ term is twice it. Retardation enters only through
$\chi_q$ and through the terms proportional to $\Delta=q^2$, which multiply the forms
$\Gq$ and $\Hq$. The transverse sums contain no term in $\Delta$: they hold as written for any
$\chi$. Only the small--small terms $\beta^2$ and $\gamma_1^2$ of the Coulomb and magnetic sums bring
in the factor $(l_0^2-1)^{-1}$.

The proof follows the route of \cite{whelan86} and is given in \ref{app:proof}.

\section{Numerical aspects and an application}
\label{sec:numerics}

\subsection{Checks}
We have compared the sum rules with direct summation over $\kappa$ and $\kappa'$ using exact Wigner
coefficients, for the kinematics of the coplanar symmetric $(e,2e)$ experiments on the K shell of
gold \cite{nakel99}: $E_0-c^2=500$\,keV, $E'-c^2=209.65$\,keV, for which $k=233.94$, $k'=136.27$,
$q=77.86$\,a.u., $\chi=1.14962$ and $\chi_q=1.05453$. Table~\ref{tab:check} shows the retarded
Coulomb and magnetic sums. With 40-digit arithmetic the direct sums and the sum rules agree to
$10^{-38}$, both in the static and in the retarded case, for $l_0=2$ to 40. The transverse sums
(\ref{eq:SC0}) and (\ref{eq:S00}), checked against angular factors computed directly from the spinors
rather than from table~\ref{tab:ang}, agree to double precision.

\begin{table}
\caption{\label{tab:check}Direct summation compared with the retarded sum rules (\ref{eq:SC}) and
(\ref{eq:SM}) for $k=233.94$, $k'=136.27$, $q=77.86$\,a.u.}
\centering
\resizebox{\textwidth}{!}{\begin{tabular}{@{}lllll}
\br
$l_0$&$\Sigma^C$ direct&$\Sigma^C$, (\ref{eq:SC})&$\Sigma^M$ direct&$\Sigma^M$, (\ref{eq:SM})\\
\mr
2 &$6.294337133680\times10^{-1}$&$6.294337133680\times10^{-1}$&$1.876729211839\times10^{-1}$&$1.876729211839\times10^{-1}$\\
5 &$8.046260906507\times10^{-2}$&$8.046260906507\times10^{-2}$&$1.766074810820\times10^{-2}$&$1.766074810820\times10^{-2}$\\
10&$2.917614082758\times10^{-3}$&$2.917614082758\times10^{-3}$&$5.698482909463\times10^{-4}$&$5.698482909463\times10^{-4}$\\
20&$4.009470509746\times10^{-6}$&$4.009470509746\times10^{-6}$&$7.300269688317\times10^{-7}$&$7.300269688317\times10^{-7}$\\
\br
\end{tabular}}
\end{table}

\subsection{Convergence of the partial-wave sums}
\begin{figure}
\centering
\includegraphics[width=0.55\textwidth]{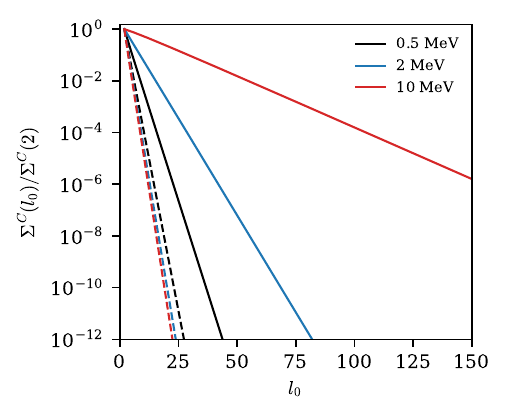}
\caption{\label{fig:tail}Fraction $\Sigma^C(l_0)/\Sigma^C(2)$ of the Coulomb dipole Bethe sum
contained in partial waves $l\ge l_0$, for impact energies of 0.5, 2 and 10\,MeV and scattered
energies of 209.65, 900 and 4900\,keV. Full curves: retarded ($\chi_q$); dashed curves: static
($\chi$).}
\end{figure}

The decisive quantity is $\eta=\cosh^{-1}\chi_q$. In the static case $\chi$ depends only on the
ratio $k'/k$, and the partial-wave tail is essentially independent of the impact energy. With
retardation,
\begin{equation}
\chi_q-1=\frac{(k-k')^2-q^2}{2kk'},
\end{equation}
and since $c(k-k')-\omega\to0$ as both particles become ultrarelativistic, $\chi_q\to1$, $\eta\to0$
and the tail lengthens. Figure~\ref{fig:tail} shows $\Sigma^C(l_0)/\Sigma^C(2)$ for an energy loss
of about 58\% of the kinetic energy at three impact energies. The number of partial waves needed for
the tail to fall below $10^{-6}$ of the total is 15, 13 and 13 in the static case at 0.5, 2 and
10\,MeV, but 23, 43 and 156 with retardation ($\chi_q=1.0545$, 1.0150 and 1.0011). In a
distorted-wave calculation the exact partial waves must be retained up to the point where the Bethe
approximation becomes accurate; beyond it the sum rules give the whole remainder at the cost of two
Legendre functions, and their advantage grows with the energy.

\section{Discussion}
\label{sec:discussion}

The simplification found for neutral targets in the nonrelativistic case \cite{whelan86} survives
the transition to Dirac waves and to the retarded M{\o}ller interaction: the relativistic Bethe
matrix elements of every multipole order are finite combinations of Legendre functions of the
second kind of a single argument, and for dipole transitions the sums over all high partial waves
collapse to the values of $Q_{l_0-1}$ and $Q_{l_0}$. Retardation enters only through the argument
$\chi_q$ and through terms proportional to $q^2$. Combined with the backward recurrence for $Q_l$
this provides a simple and accurate completion of partial-wave sums in relativistic distorted-wave
calculations: the exact distorted waves are used up to $l_0$ and the remainder is added from
(\ref{eq:SC})--(\ref{eq:S00}).

Three remarks are in order. First, the results have been given in the Feynman gauge, which is the
form of the M{\o}ller interaction used in the rDWBA \cite{keller94a,nakel99}. In the Coulomb gauge
the Coulomb term is instantaneous and the static Coulomb sum rule applies to it exactly, while the
transverse interaction takes a different multipole form; the Bethe integrals are again of the type
covered by the theorem of section~\ref{sec:retarded}. Second, the $\lambda=2$ transverse term,
whose kernel falls off as $r^{-3}$, has a rapidly convergent partial-wave sum and is evaluated
directly from (\ref{eq:theorem}). Third, the theorem holds for every multipole order, so the method
of \cite{whelan86} for constructing sum rules can be applied to higher multipoles in the same way;
we have not pursued this here, since the higher multipoles do not limit the convergence.

\appendix
\section{Proof of the sum rules}
\label{app:proof}

Let $t(l)$ denote the contribution of a single incident orbital angular momentum $l$ to one of
the sums $\Sigma^X(l_0)$. Using (\ref{eq:I0dip}) and (\ref{eq:I0mono}) with $\chi$ arbitrary, each
$t(l)$ is a quadratic form in $Q_{l-2},\dots,Q_{l+2}$, which the recurrence (\ref{eq:rec}) reduces to a
quadratic form in $Q_{l-1}$ and $Q_l$ with coefficients rational in $l$, $\chi$, $k$ and $k'$. We seek
$F(l)=f_0(l)Q_{l-1}^2+f_1(l)Q_{l-1}Q_l+f_2(l)Q_l^2$ such that
\begin{equation}
t(l)=F(l)-F(l+1)
\label{eq:tele}
\end{equation}
identically. Writing $Q_{l+1}$ in $F(l+1)$ by means of (\ref{eq:rec}) and equating the coefficients
of $Q_{l-1}^2$, $Q_{l-1}Q_l$ and $Q_l^2$ gives three linear difference equations for $f_0$, $f_1$ and
$f_2$. Taking the $f_i$ to be polynomials in $l$, divided by $l^2-1$ for the small--small terms,
these equations reduce to a linear system for the polynomial coefficients, whose solution is unique
and is given in section~\ref{sec:sumrules}. Since $Q_l(\chi)$ decreases like
$\rme^{-(l+\frac12)\eta}l^{-1/2}$ while the $f_i$ grow only polynomially, $F(L)\to0$ as $L\to\infty$,
and the sum telescopes: $\Sigma(l_0)=\sum_{l\ge l_0}[F(l)-F(l+1)]=F(l_0)$. The identity
(\ref{eq:tele}) holds for arbitrary $\chi$, $k$ and $k'$, which is why the same formulae apply in
the static and retarded cases; the relation between $\chi$, $k$ and $k'$ enters only through
$\Delta$, and $\Delta=0$ reduces the expressions to the static forms. For the nonrelativistic term
this reproduces the derivation of \cite{whelan86}, where the static relation was used from the
outset.

\section*{Data availability statement}
All data that support the findings of this study are included within the article. The computer
codes used to generate the numerical results are available from the author upon reasonable request.

\section*{References}
\bibliographystyle{unsrt}
\bibliography{sumrule}

\begin{thebibliography}{10}

\bibitem{seaton61}
M.~J. Seaton.
\newblock The evaluation of partial wave integrals in the {B}orn approximation.
\newblock {\em Proc. Phys. Soc.}, 77:184, 1961.

\bibitem{burgess70}
A.~Burgess, D.~G. Hummer, and J.~A. Tully.
\newblock Electron impact excitation of positive ions.
\newblock {\em Phil. Trans. R. Soc. A}, 266:225, 1970.

\bibitem{burgess74}
A.~Burgess.
\newblock Coulomb integrals: tables and sum rules.
\newblock {\em J. Phys. B: At. Mol. Phys.}, 7:L364, 1974.

\bibitem{biedenharn56}
L.~C. Biedenharn.
\newblock A note on {S}ommerfeld's bremsstrahlung formula.
\newblock {\em Phys. Rev.}, 102:262, 1956.

\bibitem{burgess74a}
A.~Burgess and V.~B. Sheorey.
\newblock Electron impact excitation of the resonance lines of alkali-like
  positive ions.
\newblock {\em J. Phys. B: At. Mol. Phys.}, 7:2403, 1974.

\bibitem{whelan86}
Colm~T. Whelan.
\newblock On the {B}ethe approximation to the reactance matrix.
\newblock {\em J. Phys. B: At. Mol. Phys.}, 19:2343, 1986.

\bibitem{burgess87}
Alan Burgess and Colm~T. Whelan.
\newblock {BETRT}: a procedure to evaluate the cross section for
  electron--hydrogen collisions in the {B}ethe approximation to the reactance
  matrix.
\newblock {\em Comput. Phys. Commun.}, 47:295--304, 1987.

\bibitem{keller94a}
S.~Keller, Colm~T. Whelan, H.-J. Ast, H.~R.~J. Walters, and R.~M. Dreizler.
\newblock Relativistic distorted-wave {B}orn calculations for (e,2e) processes
  on inner shells of heavy atoms.
\newblock {\em Phys. Rev. A}, 50:3865, 1994.

\bibitem{nakel99}
W.~Nakel and Colm~T. Whelan.
\newblock Relativistic (e,2e) processes.
\newblock {\em Phys. Rep.}, 315:409, 1999.

\bibitem{qiu07}
Cheng-Wei Qiu, Le-Wei Li, Sa{\"\i}d Zouhdi, Tat-Soon Yeo, and Qun Wu.
\newblock On the integral identities consisting of two spherical {B}essel
  functions.
\newblock {\em IEEE Trans. Antennas Propag.}, 55:240--244, 2007.

\bibitem{grant07}
I.~P. Grant.
\newblock {\em Relativistic Quantum Theory of Atoms and Molecules}.
\newblock Springer, New York, 2007.

\bibitem{johnson07}
W.~R. Johnson.
\newblock {\em Atomic Structure Theory: Lectures on Atomic Physics}.
\newblock Springer, Berlin, 2007.

\bibitem{watson44}
G.~N. Watson.
\newblock {\em A Treatise on the Theory of {B}essel Functions}.
\newblock Cambridge University Press, Cambridge, 2nd edition, 1944.

\end{thebibliography}

\end{document}